\documentclass[reprint,aip,jap,amsmath,amssymb,nofootinbib]{revtex4-2}
\usepackage{amssymb}
\usepackage{lipsum}
\usepackage{amssymb}
\usepackage{bbold}
\usepackage{amssymb}
\usepackage{graphicx}
\usepackage{graphics}
\usepackage{amsmath}
\usepackage{placeins}
\usepackage{hyperref}
\usepackage[dvipsnames]{xcolor}
\usepackage{wasysym}
\usepackage{soul}
\usepackage{float}
\usepackage{array}
\usepackage{tabulary}
\usepackage{caption}
\usepackage{fancyhdr}
\usepackage{wrapfig}

\begin{document}

\title{Useful Circuit Analogies to Model THz Field Effect Transistors}

\author{Adam Gleichman}
\affiliation{Department of Electrical and Computer Engineering, Michigan State University, 428 S. Shaw Ln., East
Lansing, MI 48824, USA}

\author{Kindred Griffis}
\affiliation{Department of Electrical and Computer Engineering, Michigan State University, 428 S. Shaw Ln., East
Lansing, MI 48824, USA}

\author{Sergey V. Baryshev}
\email{serbar@msu.edu}
\affiliation{Department of Electrical and Computer Engineering, Michigan State University, 428 S. Shaw Ln., East
Lansing, MI 48824, USA}
\affiliation{Department of Chemical Engineering and Materials Science, Michigan State University, 428 S. Shaw Ln., East
Lansing, MI 48824, USA}

\begin{abstract}
    The electron fluid model in plasmonic field effect transistor (FET) operation is related to the behavior of a radio-frequency (RF) cavity. This new understanding led to finding the relationships between physical device parameters and equivalent circuit components in traditional parallel resistor, inductor, and capacitor (RLC) and transmission models for cavity structures. Verification of these models is performed using PSpice to simulate the frequency dependent voltage output and compare with analytical equations for the drain potential as a function of frequency. 
\end{abstract}

\maketitle

\section{Introduction}

Moore's law predicted that every 2 years that the transistor count would double for a constant amount of area real estate, power consumption, and would cost the same\cite{moore2006cramming}\cite{shalf2020future}.
This progression of higher availability of transistors in integrated circuits was fostered by discoveries like lithography.
These improvements to the amount of transistors at lower cost led to massive windfalls for growth in automated control systems, data processing, portable communication systems, and cost for broad consumer use \cite{shalf2020future}.
Moore's Law is said to end at some point in the future and projections currently are around 2025. \cite{shalf2020future}
Although in the beginning when Moore first made his prophecy it was initially thought to last for only 20 years.\cite{moore2006cramming}\cite{shalf2020future}
Many advancements did delay the slowing down of transistor development, but ultimately there are limitations of material properties and the physics of energy transportation. \cite{shalf2020future} 
This situation leads to a question of "what is the next technological advancement for computing devices?" 
Microwave devices could possibly be the next advancement in the race for better transistors by using the physics of the device to create better performance. 

Dyakonov and Shur proposed that field-effect transistors (FETs) can have a new operation mode that utilizes a relationship between radiation at the gate and steady current across the channel that creates Langmuir waves in the channel \cite{dyakonov_shallow_1993,dyakonov_detection_1996}. 
They describe an instability under the condition where a short channel FET operates with a large amount of electrons travelling in the channel, which leads to many electron on electron collisions\cite{dyakonov_shallow_1993}. 
This system is modelled as a 2D gas which describes instability that exists for electrons that are slower than their own saturation velocity\cite{dyakonov_shallow_1993}\cite{dyakonov_choking_1995}.
High concentration of electrons in the channel leads to many electron on electron collisions and fluid choking will occur at the boundary where the channel meets the drain \cite{dyakonov_shallow_1993}\cite{dyakonov_choking_1995}.
The electrons accelerate across the channel in subsonic flow until their velocity reaches the speed of sound at the boundary of the drain \cite{dyakonov_choking_1995}. 
This constant electron velocity present at the drain means that from the outside observer perspective, the drain has constant DC current \cite{dyakonov_choking_1995}.
The high concentration of electrons slow their group velocity to be less than their saturation velocity, but the Langmuir waves phase velocity can exceed the electron transit time \cite{dyakonov_shallow_1993}. 
Using the phase velocity of Langmuir waves instead of the group velocity of the electrons leads to terahertz level radiation.
This output terahertz frequency at the gate can used in place of the traditional switching speed for FETs.
This instability relationship is dependent on the volume of the channel structure\cite{dyakonov_shallow_1993}. 
The DC bias potential controls the depth of the channel, which controls the excitation frequency of the channel to create current across the channel. \cite{dyakonov_shallow_1993}\cite{dyakonov_detection_1996}

\begin{figure}[h]
	\centering\includegraphics[width=7cm]{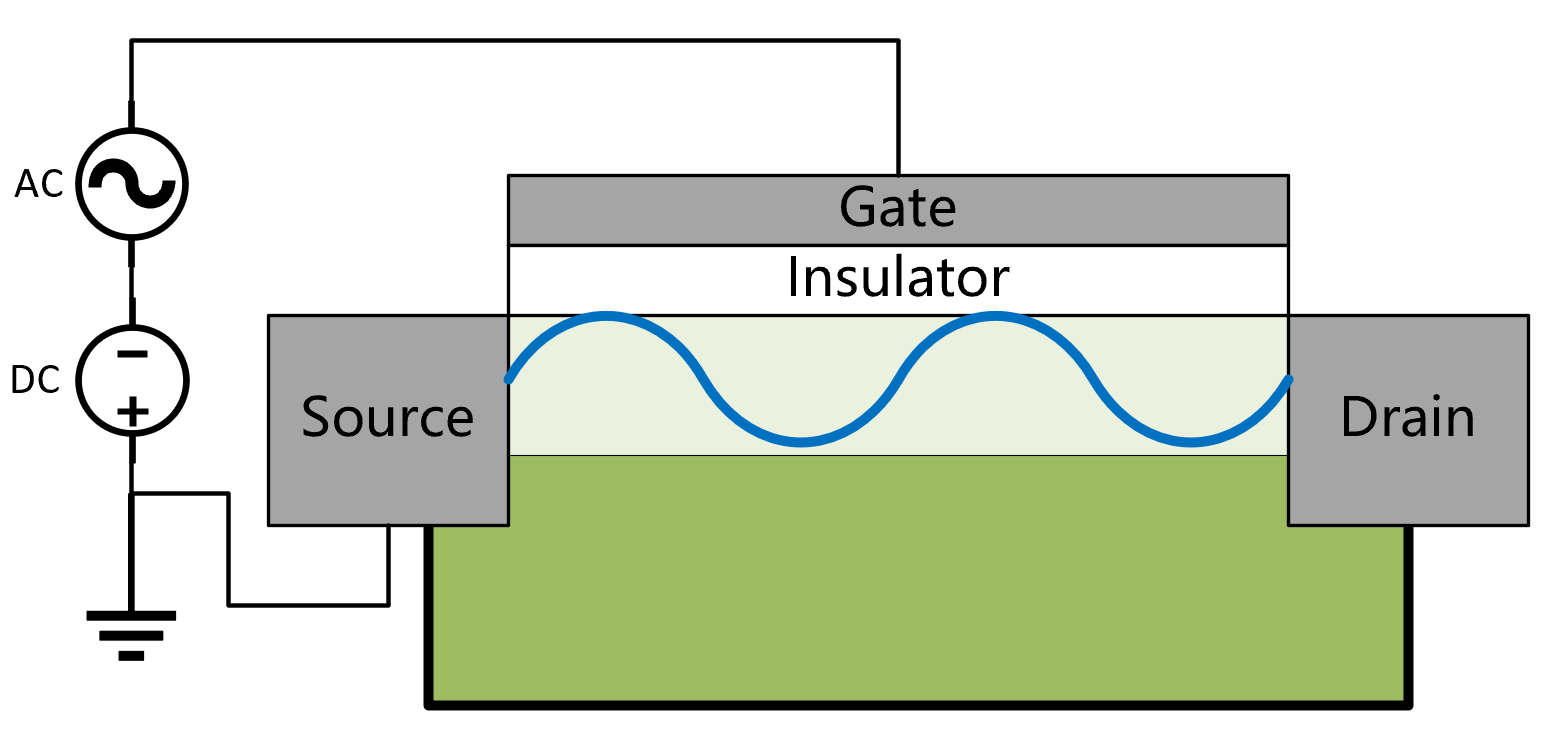}
	\caption{A Plasmonic FET operating as a detector.}
        \label{F:1}
\vspace*{0mm}
\end{figure}

Creating models of plasmonic operating devices is difficult.
Previous models\cite{liu_compact_2018,liu_compact_2019,monster2} were designed as exceptional level complexity or were dependent on empirical data \cite{cheng_mosfet_2005}. In Ref.[\onlinecite{liu_compact_2019}] a so-called MOSFET segmentation concept was introduced. Because each segment was on a nm to sub-nm length scale the model had to be solved inside EKV (Enz, Krummenacher, Vittoz) model framework, thereby making the segmentation concept dependent on dozens of free parameters (previously optimized only for classical operation of short-channel MOSFETs) and hence making difficult for practical applications. Conceptually simple and easy to interpret THz plasma FET models, if created, could play vital role to allowing the development of architectures and ICs for future microelectronics.

In this paper, we show that plasmonic FET, behaving as a resonating device caused by standing waves that arise from the boundary conditions of the source and drain contacts, as illustrated in Figure~\ref{F:1}, can be treated as a classical radiofrequency (RF) cavitiy. Small signal parallel RLC circuit model for solving first resonant mode and transmission line model capturing higher order modes are introduced and shown to have excellent agreement with analytical model for Si THz FET operation. Finally, simple symbolic PSpice codes are developed and presented in the Appendices at the end of this paper. All PSpice parameters are physical and can be directly calculated using geometry and material properties of the channel and gate.

\section{Theories and Equations}

The signal between the gate and the source is a combination of an AC signal and DC bias potential shown in Figure~\ref{F:1} \cite{dyakonov_detection_1996}. 

\begin{figure}[h]
	\centering\includegraphics[width=7cm]{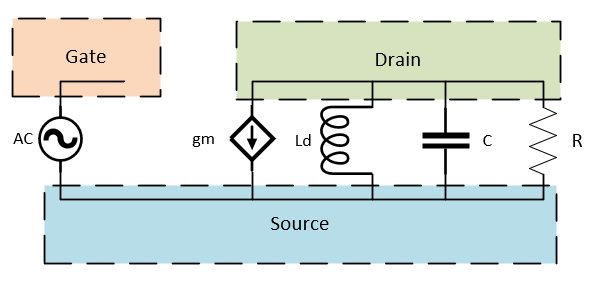}
	\caption{RLC Lumped Model.}\label{F:2}
\vspace*{0mm}
\end{figure}

In Figure~\ref{F:2}, a voltage controlled current source (VCCS) with the gate's resistance controls the gain of the device at the drain. VCCS need a value for the transconductance in PSpice which $Gain = g_m \cdot R_g$. 
To find the transconductance, the start would be the equation for current at the drain which would be

\begin{equation}
    i_D = K(V_{AC} + V_{DC} - V_{T})^2. 
\end{equation}

Which is then expressed in terms of the DC potential between the gate and source, $V_{DC}$, and the threshold potential, $V_{T}$, as
\begin{equation}
    i_D = K(V_{DC} - V_{T})^2 + 2KV_{AC}[V_{DC} - V_{T} ] + KV^2_{AC}.
\end{equation}

The FET amplifies the AC signal due to amplification from the standing wave mechanism in the device \cite{dyakonov_generation_2010}\cite{dyakonov_shallow_1993}. The DC bias only matters with the depth of the depletion channel. Which means that we can neglect the purely DC term in Equation 2 \cite{dyakonov_shallow_1993}.
Transconductance is defined in terms of the gain constant, $K$, which allows for solving the system in terms of circuit components instead of arbitrary gain, which is given by

\begin{equation}
    g_m = 2K(V_{DC} - V_{T}).
\end{equation}

The FET in \cite{dyakonov_detection_1996}, the drain is disconnected from the source which for the simplicity of the case we treat as the source is grounded, demonstrated in Figure~\ref{F:2}. 
For practical purposes, there will be some DC potential at the source usually, but that will also be applied to the gate which means that the potential difference between the gate and source is constant. 
Since an open circuit is placed in the drain, the drain current equation in (2) needs to be adjusted because all DC terms are now 0. 
So 
\begin{equation}
    i_D = KV_{AC}^2 = \frac{g_m}{2(V_{DC} - V_{T})}V_{AC}^2
\end{equation}

Equation (4) allows for direct evaluation of the potential difference between the drain and the source with multiplying Equation (4) by the resistance of the device channel. 
Also using the root mean square for the AC potential from the source

\begin{equation}
    \Delta V = \frac{V_{AC}^2}{4(V_{DC} - V_{T})}g_m R
\end{equation}

The VCCS with a parallel resistor matches the gain of the FET detector. 
The gain is in terms of the input AC source amplitude, which allows PSpice to correctly model the system in Equation (5). 

Assuming the entire channel is gated, the total capacitance in the channel is a series connection between the capacitance due to the gate insulator and the capacitance in the channel.
First, the capacitance from the insulation at the gate is given by \cite{ytterdal2003device}
\begin{equation}
    C_i = \frac{LW\epsilon_0\epsilon_{I}}{t_{ox}}.
\end{equation} 
The capacitance due to the depletion in channel of the FET is dependent on the depletion distance needs to be solved for \cite{ytterdal2003device}

\begin{equation}
    d_d = \sqrt{\frac{2\epsilon_0\epsilon_S \psi_S}{qN_b}}.
\end{equation}
$\psi_S$ is the surface potential applied to the channel defined as \cite{ytterdal2003device}
\begin{equation}
    \psi_S = 2V_{TH} \ln{\frac{N_b}{n_i}}.
\end{equation}

Where $\epsilon_S$ is the relative permittivity of the substrate, $q$ is the charge of the electron (in Coulombs, $\mbox{C}$), and $N_b$ is the concentration of donor electrons in the silicon channel. And $n_i$ is the intrinsic concentration of the channel. Both these concentration values are defined in per volume units usually given in $\mbox{cm}^{-3}$. Equation 8 and 13 need the thermal potential of the transistor, $V_{TH}$, which is solved by 
\begin{equation}
    V_{TH} = \frac{kT}{q}.
\end{equation}
Where $T$ is temperature in Kelvin, $k$ is the Boltzman constant, and $q$ will be the charge of an electron. \\

The depletion channel capacitance is therefore \cite{ytterdal2003device}
\begin{equation}
    C_d = \frac{LW\epsilon_0\epsilon_{S}}{d_d}.
\end{equation}
And the total capacitance, $C_{tot}$, for the channel is solved by taking the parallel combination of the capacitance of the depletion region, $C_d$, with the capacitance of the insulator from the gate, $C_i$,\cite{ytterdal2003device}
\begin{equation}
    C_{tot} = \frac{C_iC_d}{C_i + C_d}.
\end{equation}

Calculation of the Drude inductance is performed by solving for the electron sheet density in terms of the ideality constant of the FET, $\eta$ \cite{ytterdal2003device}
\begin{equation}
    \eta = 1 + C_d / C_i.
\end{equation}

The expression of the initial sheet electron is given by \cite{ytterdal2003device}

\begin{equation}
    n_0 = \frac{\eta V_{TH} C_{OX}}{2q}.
\end{equation}

Where the capacitance of the oxide per unit area is $C_{OX} = \frac{\epsilon_I \epsilon_0}{t_{ox}}$. 
The concentration of electrons in the channel per unit area once the thermal effects are accounted for is \cite{ytterdal2003device}

\begin{equation}
    n_s = n_0 \ln{ [1 + 0.5\exp{(\frac{V_{DC} - V_{T}}{\eta V_{TH}}})]}.
\end{equation}

This concentration of electrons per unit area in the channel will be used to solve for the the Drude inductance, $L_{drude}$, because the concentration effects the amount of electron on electron collisions that are present in the system. The Drude inductance equation is as follows \cite{shur_plasma_2010}

\begin{equation}
    L_{drude} = \frac{L \cdot m_{eff} \cdot m_0}{q^2\alpha^2 n_s W}.
\end{equation}

From here, the resistance is calculated in order is calculated in order to model the leakage between the gate and the drain the quality factor equation as a relationship between the fundamental mode, the total capacitance of the channel, and the quality factor $R=\frac{Q}{\omega_0 C_{total}}$. This is important to make sure that the quality factor of the new models will still match results in \cite{liu_compact_2018} because the bandwidth of the system is dependent on the quality factor from $Q=BW^{-1}$.

The transconductance equation for the model \cite{sze2021physics},
\begin{equation}
    g_m = \frac{W\mu_n}{L}C_{OX}(V_{DC} - V_{T}) ,
\end{equation}

\section{Model Validation}

\begin{figure}[h]
	\centering\includegraphics[width=7cm]{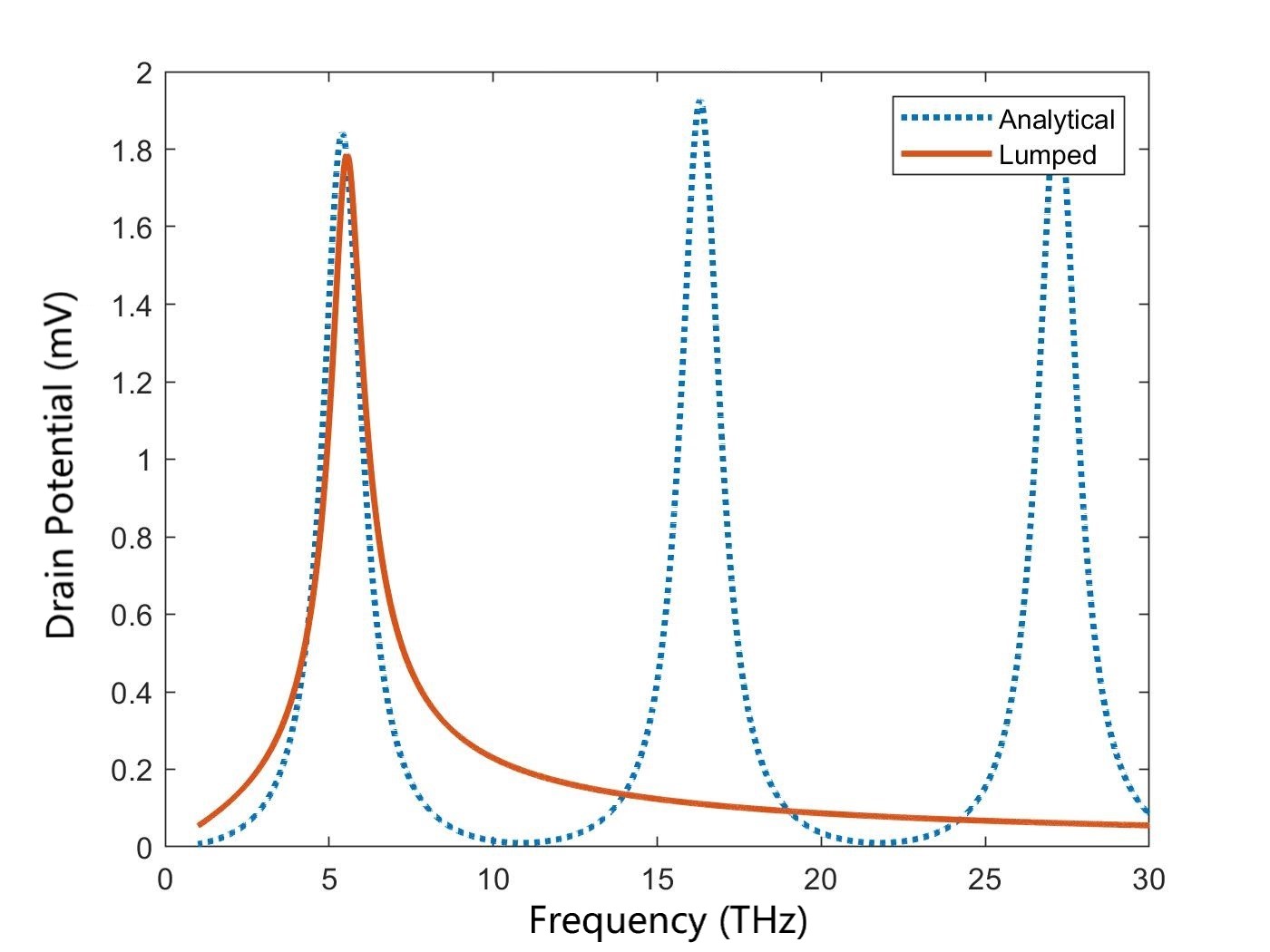}
	\caption{The lumped RLC model in appendix 1 (solid line) compared to the analytical model (dotted line).}\label{F:3}
\vspace*{0mm}
\end{figure}

\begin{figure}[h]
	\centering\includegraphics[width=7cm]{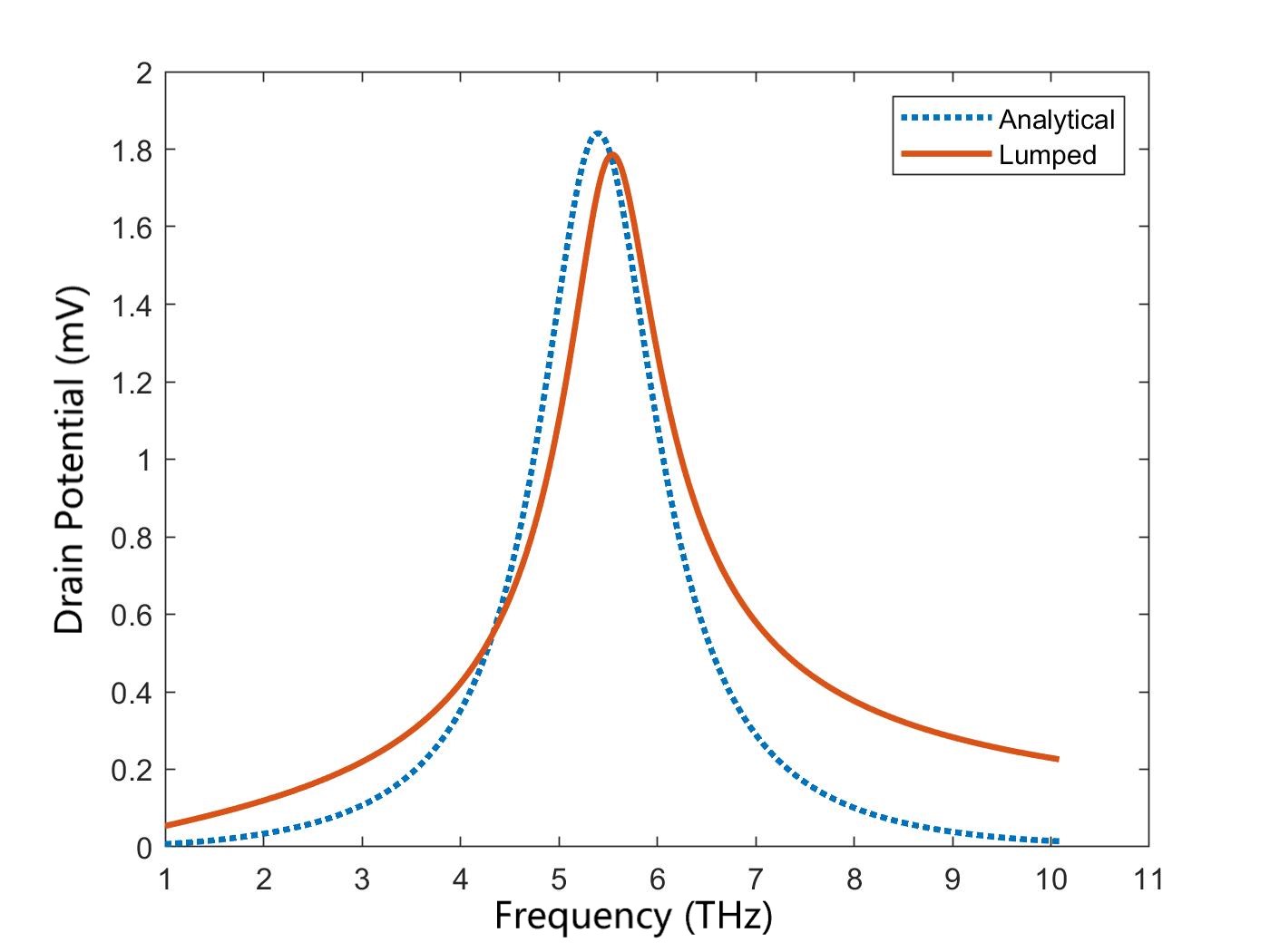}
	\caption{Close up of the fundamental mode of the lumped RLC in appendix 1 (solid line) and the analytical model (dotted line).}\label{F:4}
\vspace*{0mm}
\end{figure}

\begin{figure}[h]
	\centering\includegraphics[width=7cm]{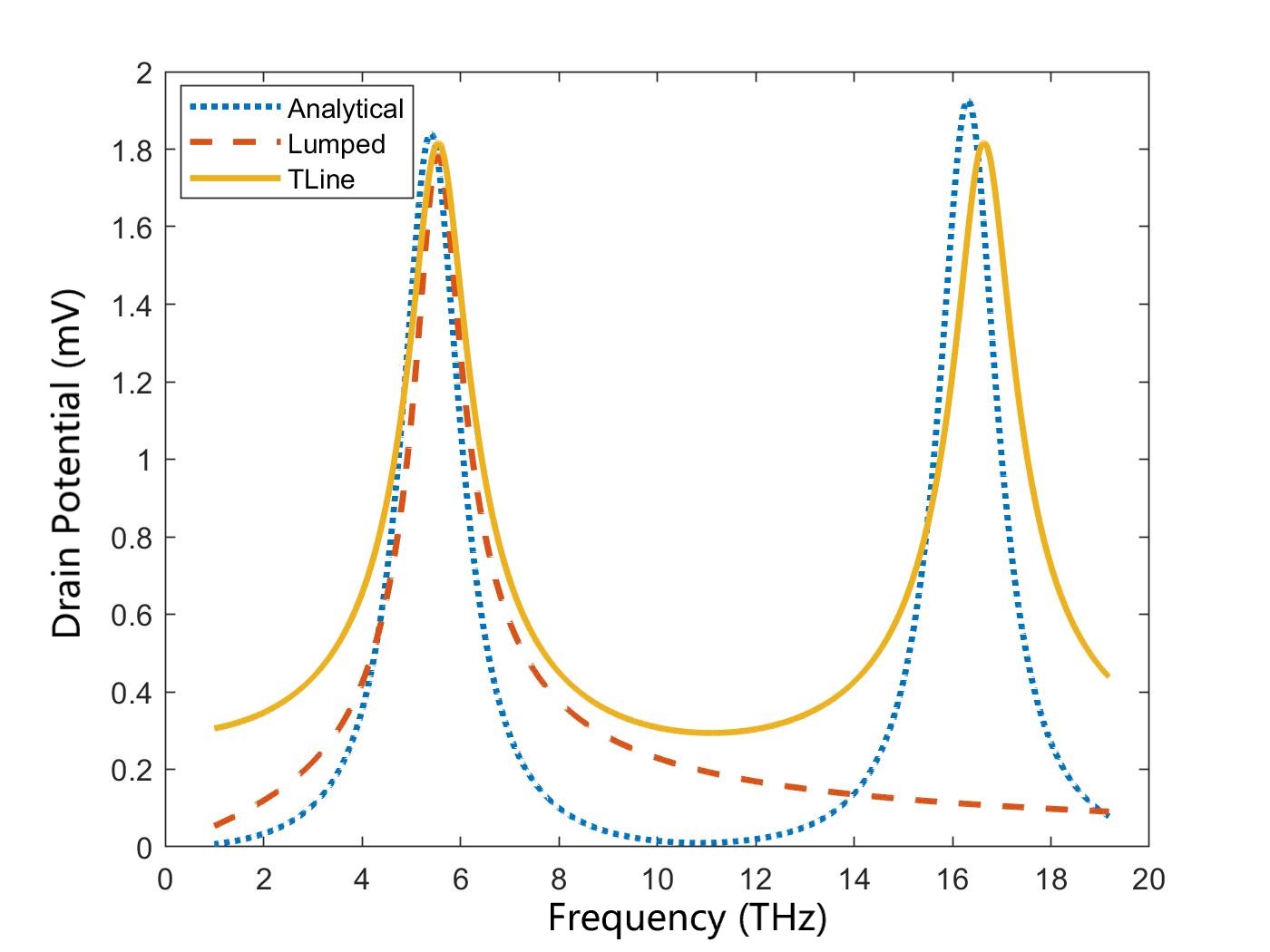}
	\caption{Comparison of the lumped RLC in appendix 1 (dash line), transmission line in appendix 2 (solid line), and analytical model (dotted line).}\label{F:6}
\vspace*{-4mm}
\end{figure}

To validate agreement between the lumped and transmission line models match and the fluid plasmonic model, we considered a silicon channel ($\epsilon_S=11.9$ from \cite{ioffe}) with a 3D concentration of $N_b=10 \times 10^{17} \: \mbox{cm}^{-3}$ and an intrinsic concentration of silicon of $n_i=10^{10} \mbox{cm}^{-3}$ from \cite{ioffe} with a silicon oxide insulator ($\epsilon_I=3.9$ from \cite{ioffe}) that has a thickness of $t_{ox}=4.315 \; \mbox{nm}$. 
The mobility of the substrate is $\mu = 0.1 \: \frac{\mbox{m}^2}{\mbox{V} \cdot \mbox{s}}$ and the effective mass used is 0.19 (from \cite{ioffe}). 
Dimensions of the device are a length of $L=25 \: \mbox{nm}$ and the width of $W=5 \: \mu \mbox{m}$. 
A DC potential is applied between the gate and source of $V_{DC}=0.6 \: \mbox{V}$ with a thermal potential of $V_{TH}=0.28 \: \mbox{V}$ and the applied AC signal amplitude is $V_{AC}=100 \; \mbox{mV}$.
All calculations were made with an assumption of room temperature operation at $T=300\; \mbox{K}$.
The RLC solution to these system was a transconductance of $g_m=12.7 \: \mbox{mS}$, $L_{drude}=8.352 \; \mbox{pH}$, $C_{tot}=9.864659\times 10^{-17} \: \mbox{F}$, and a resistance of 1800 $\Omega$.
The corresponding PSpice file for the RLC simulation is included in appendix 1.
The parallel RLC model is simulated in PSpice, which exported the data of the drain-source potential ($\Delta V$) as a function of frequency (with respect to $V_{AC}$) into a comma-separated values format for MATLAB to import. 
MATLAB is used to compare the PSpice models with the fluid model as shown in Figures~\ref{F:3},~\ref{F:4}, and~\ref{F:6}.

The parallel RLC model is successful in replicating the fundamental resonant mode, but fails to create the higher order modes from the fluid model.
Lack of the higher order modes present motivates the use of a lossless transmission line component in PSpice, Figure~\ref{F:3} is the circuit model used (file is in Appendix 1), to generate those higher order modes.

\begin{figure}[h]
	\centering\includegraphics[width=7cm]{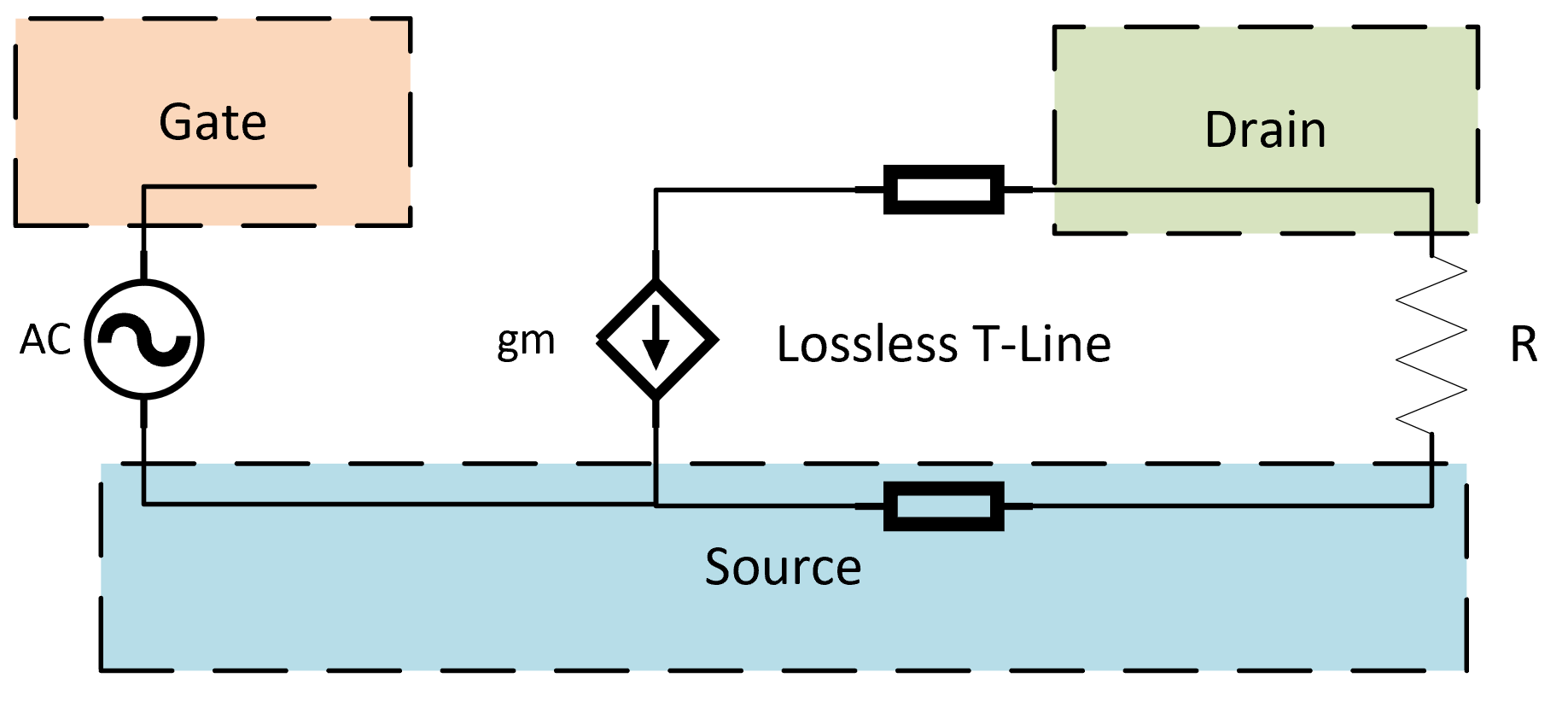}
	\caption{Transmission Line Model}\label{F:5}
\vspace*{0mm}
\end{figure}

A lossless half wave transmission line is shown to replicate the fundamental modes with the higher order modes \cite{pozar2011microwave}. 
Usually, an open circuited half wave transmission line is equivalent to a passive parallel RLC model, however this transmission line model requires the same resistor connected as a load so that the gain from the transconductance is still accounted for in the model \cite{pozar2011microwave}. 

\begin{center}
\begin{tabular}{ c|c|c|c }
 \textbf{Name} & \textbf{Symbol} & \textbf{Value} & \textbf{Units} \\
 \hline
 \hline
 3D Concentration of Donors & $N_b$ & $10\times10^{17}$ & $\mbox{cm}^{-3}$ \\  
 \hline
 Intrinsic Concentration of Si & $n_i$ & $1\times10^{10}$ & $\mbox{cm}^{-3}$
 \\
 \hline
 Channel Length & $L$ & 25 & $\mbox{nm}$ \\
 \hline
 Channel Width & $W$ & 5 & $\mu \mbox{m}$ \\
 \hline
 Insulator Thickness & $t_{ox}$ & 4.315 & $\mbox{nm}$ \\
 \hline
 Mobility of the Channel & $\mu$ & 0.1 & $\frac{\mbox{m}^2}{\mbox{V}\cdot \mbox{s}}$ \\
 \hline
 Threshold Potential & $V_{T}$ & 0.28 & $\mbox{V}$ \\
 \hline
 DC Bias Potential & $V_{DC}$ & 0.6 & $\mbox{V}$ \\
 \hline
 Effective Mass of Si & $m_{eff}$ & 0.19 & -- \\
 \hline
 Relative Permittivity of Si & $\epsilon_S$ & 11.9 & $\mbox{n/a}$ \\
 \hline
 Relative Permittivity of SiO2 & $\epsilon_I$ & 3.9 & $\mbox{n/a}$ \\
 \hline
 Temperature & $T$ & 300 & K \\

\end{tabular}
\end{center}

In Figure~\ref{F:6}, the result of the transmission line model (file is in Appendix 2) has good agreement with the fundamental model by having a consistent quality factor, but with the higher order modes that are not present in the parallel RLC model in Figures~\ref{F:3} and~\ref{F:4}. This transmission line model increasingly shifts the center frequency of  higher order modes, this is also dependent on the transconductance of the VCCS. 

\section{Conclusion}
The results of this paper show that cavity-inspired circuit models demonstrate promising results for small signal/ ultra high frequency design. Our results illustrate that the cavity behavior in plasmonic FET operation leads to simple and effective circuit models for small signal evaluations with only five parameters solved from physical dimensions. 

\section{Appendix 1: Parallel RLC .cir File}

\noindent PARALLEL RLC \\
$* V_{in} = V^2_{AC} / (4 * (V_{DC} - V_T) ), V_{AC} = 0.01~\text{V}, V_{DC} - V_T = 0.32~\text{V} $\\
\mbox{Vin}\; 1\; 0\; \mbox{AC}\; 7.8125e-5\\
* \mbox{This is the voltage dependent current source} \\
\mbox{G1}\; 3\; 0\; 1\; 0\; 12.7\mbox{m} \\
* \mbox{This is the Drude inductance} \\
\mbox{L1}\; 3\; 0\; 8.352\mbox{e}-12 \\
* \mbox{This is the resistance}\\
\mbox{R1}\; 3\; 0\; 1800 \\
* \mbox{This is the total capacitance of the channel} \\
\mbox{C1}\; 3\; 0\; 9.86465905084\mbox{e}-17 \\
\\
* \mbox{This creates a sweep of 5000 points from 1 to 30 THz} \\
\mbox{.AC}\; \mbox{LIN}\; 5000\; 1\mbox{T}\; 30\mbox{T} \\
\mbox{.PROBE} \\
\mbox{.OP} \\
\mbox{.END} \\

author's note: Some of the gain scaling was multiplied to the AC input potential instead of the transconductance term because solving a number and adjusting the input voltage was easier in PSPICE than trying to put transconductance as a multiple of the AC voltage divided by the DC potential. This is also done in the transmission line file. The corresponding figure to this model layout is Figure~\ref{F:2}.

\section{Appendix 2: Transmission Line .cir File}

\noindent TRANSMISSION LINE MODEL \\
$* V_{in} = V^2_{AC} / (4 * (V_{DC} - V_T) ), V_{AC} = 0.01~\text{V}, V_{DC} - V_T = 0.32~\text{V} $\\
\mbox{Vin}\; 1\; 0\; \mbox{AC}\; 7.8125e-5 \\
* \mbox{This is the voltage dependent current source} \\
\mbox{G1}\; 2\; 0\; 1\; 0\; 0.012923082392042 \\
* This is the transmission line element. Impedance and center frequency are solved from lumped RLC components. \\
\mbox{T1}\; 2\; 0\; 4\; 0\; \mbox{Z0} = 290.974\; \mbox{f} = 5.544774\mbox{THZ} \\
* \mbox{This is the load resistance}\\
\mbox{RL}\; 4\; 0\; 1800 \\
\\
* \mbox{This creates a sweep of 5000 points from 1 to 30 THz} \\
\mbox{.AC}\; \mbox{LIN}\; 5000\; 1\mbox{T}\; 30\mbox{T} \\
\mbox{.PROBE} \\
\mbox{.OP} \\
\mbox{.END} \\

author's note: Some of the gain scaling was multiplied to the AC input potential instead of the transconductance term because solving a number and adjusting the input voltage was easier in PSPICE than trying to put transconductance as a multiple of the AC voltage divided by the DC potential. This is also done in the parallel RLC line file. The corresponding figure to this model layout is Figure~\ref{F:5}.
\\
\bibliographystyle{IEEEtran}
\bibliography{TeraFET_citations}

\end{document}